\documentclass[twocolumn]{aastex63}
\usepackage{xcolor}
\usepackage{gensymb}
\usepackage[nointegrals]{wasysym}
\usepackage{amsmath}
\usepackage{verbatim}

\usepackage{threeparttable}
\usepackage{bm}

\LTcapwidth=\textwidth

\shortauthors{Kunimoto \& Bryson}
\shorttitle{Comparing ABC}

\begin{document}

\title{Comparing Approximate Bayesian Computation with the Poisson-Likelihood Method for Exoplanet Occurrence Rates}

\correspondingauthor{Michelle Kunimoto}
\email{mkunimoto@phas.ubc.ca}

\author[0000-0001-9269-8060]{Michelle Kunimoto}
\affiliation{Department of Physics and Astronomy, University of British Columbia, 6224 Agricultural Road, Vancouver, BC V6T 1Z1, Canada}

\author[0000-0003-0081-1797]{Steve Bryson}
\affiliation{NASA Ames Research Center, Moffett Field, CA 94901}

\begin{abstract}
We present \textit{Kepler} exoplanet occurrence rates inferred with approximate Bayesian computation (ABC). By using the same planet catalogue, stellar sample, and characterization of completeness and reliability as \citet{bry20}, we are able to provide the first direct comparison of results from ABC to those derived with the popular Poisson-likelihood method. For planets with orbital periods between 50 and 400 days and radii between 0.75 and 2.5 $R_{\oplus}$, we find an integrated occurrence rate $F_{0} = 0.596_{-0.099}^{+0.092}$ planets per GK dwarf star. After correcting for reliability against astrophysical false positives and false alarms, we find $F_{0} = 0.421_{-0.072}^{+0.086}$. Our findings agree within 1$\sigma$ of \citet{bry20}, indicating that the results are robust and not method-dependent.
\end{abstract}

\section{Introduction}
Exoplanet occurrence rates are fundamental observational results from exoplanet surveys, providing important constraints on planet formation and evolution theories. However, different methods for inferring occurrence rates can produce a wide range of results, and the extent to which results are model-dependent is not yet well understood.

Here, we use approximate Bayesian computation (ABC) to infer a parametric \textit{Kepler} exoplanet occurrence rate density without the need for a likelihood function. ABC has only recently been adopted by the exoplanet community, and has been applied to both discrete, grid-based \citep{hsu18, hsu19, kun20} and parametric \citep{he19} models of the planet population. Meanwhile, the Poisson-likelihood method, first introduced in \citet{you11}, is one of the most common techniques for fitting planet distribution functions to exoplanet survey data, but requires the assumption of a specific form of the likelihood. By comparing results from ABC with those from the Poisson-likelihood method, we can start to probe the consequences of this assumption.

\section{ABC Methodology}
In regular Bayesian inference, the posterior probability that a model describes the observed data is derived from our prior information about the model parameters, and the likelihood of observing the data given the model. This is appropriate when an exact likelihood function can be known, but challenges arise when the likelihood is unknown or computationally too expensive to calculate.

ABC is an approach to Bayesian inference that bypasses the need for a likelihood, instead forward modeling the data. By generating a large number of simulations using different model parameter values and quantifying the ``distance'' between the simulated and observed datasets, we can find the parameters that best describe the data. The distribution of these parameters approximates the posterior probability distribution of regular Bayesian inference.

In this work, we apply a Population Monte Carlo ABC (PMC-ABC) algorithm which uses an adaptive importance sampling scheme to evolve the ABC posterior \citep{bea09}, as follows. Using the notation of \citet{ish15}, we begin by drawing $M$ model parameters from the prior, called ``particles,'' $\{\bm{\theta}^{i}\}$ with $i \in [1,M]$. For each $\bm{\theta}^{i}$, we generate a simulated dataset $\hat{D}^{i}$ and assess its agreement with the observed data $D$ using a vector of distance functions, $\bm{\rho}^{i} = \bm{\rho}(D, \hat{D}^{i})$. The $N$ particles giving the smallest $|\bm{\rho}|$ constitute the first ``particle system'' ($S_{t=0}$). The 75\% quantile of the distances in $S_{t=0}$ determines the distance threshold vector $\bm{\epsilon}_{t=1}$ for the next generation. For this first system, all particles are assigned equal weights.

For subsequent iterations, a parameter vector $\bm{\theta}_{\text{try}}$ is drawn from the previous particle system using importance sampling and the weights of particles. A dataset is simulated using $\bm{\theta}_{\text{try}}$, and  $\bm{\theta}_{\text{try}}$ is added to $S_{t}$ if $\bm{\rho}_{\text{try}} \leq \bm{\epsilon}_{t}$. $\bm{\theta}_{\text{try}}$ are continuously drawn until $N$ particles satisfy the distance criteria. Particles are then assigned weights according to Eqn. 3 of \citet{ish15} to facilitate the importance sampling of the next generation.

With each step, $\bm{\epsilon}_{t}$ gets smaller and is satisfied by fewer draws.  We consider the algorithm converged when a large number of draws are required for N particles to satisfy $\bm{\rho}^i \leq \bm{\epsilon}_t$.

\subsection{Application to Exoplanet Occurrence Rates}

To apply ABC to exoplanet occurrence rates we need a planet population model, a way to simulate planet catalogues from the model, and distance functions to assess agreement between the simulated and observed planet catalogues.

We model the exoplanet population distribution function, $\lambda$, as a joint power law in period $P$ and radius $R_{p}$,

\begin{equation}\label{eqn:lambda}
    \lambda(P, R_{p}) = \frac{d^{2}f}{dP dR_{p}} = C P^{\beta} R_{p}^{\alpha},
\end{equation}

\noindent where $\alpha$ and $\beta$ are the power law indices and $C$ is a normalization constant such that the integral of $\lambda$ over the period and radius range of interest equals the number of planets per star, $F_{0}$:

\begin{equation}
    F_{0} = \int_{R_{p}} \int_{P} C P^{\beta} R_{p}^{\alpha} dP dR_{p}.
\end{equation}

Our simulator starts by drawing $F_{0}$, $\alpha$, and $\beta$ from prior distributions. Similar to \citet{mul18}, we then draw $N_{p} = F_{0} N_{s}$ periods and radii according to Eqn. \ref{eqn:lambda}, where $N_{s}$ is the number of stars in the sample. The periods and radii are a realization of the planet population model. We then calculate $P_{\text{det}}(P, R_{p})$, the probability that each planet both transits and would be detected by the \textit{Kepler} pipeline. We mark each planet as detected if $\text{Bernoulli}(P_{\text{det}}) = 1$. The detected planet population can then be compared to the observations.

We calculate the distance $\bm{\rho}$ between our simulated and observed planet catalogues across three dimensions: period ($\rho_{1}$), radius ($\rho_{2}$), and sample size ($\rho_{3}$). We find $\rho_{1}$ and $\rho_{2}$ using the two-sample Anderson-Darling (AD) statistic, commonly used to test whether samples are drawn from the same distribution. For sample size, we use

\begin{equation}
    \rho_{3} = \text{max}\bigg(\text{abs}\bigg(1 - \frac{l}{l_{s}}\bigg), \text{abs}\bigg(1 - \frac{l_{s}}{l}\bigg)\bigg),
\end{equation}

\noindent where $l$ and $l_{s}$ are the number of planets in the observed and simulated catalogues, respectively \citep{ish15}.

\section{Data}
We are interested in the occurrence rates of the \textit{Kepler} DR25 planet catalogue used in \citet{bry20}, consisting of planets with radii between 0.75 and 2.5 $R_{\oplus}$ and orbital periods between 50 and 400 days, orbiting a clean sample of 57,015 GK dwarf stars. We adopt their star-averaged completeness contours to calculate $P_{\text{det}}(P, R_{p})$, which take into account the geometric probability to transit, and both detection and vetting efficiencies of the DR25 pipeline. We also adopt the same uniform priors on the model parameters ($0 < F_{0} < 5$, $-5 < \alpha < 5$, and $-5 < \beta < 5$).

\section{Results}

We use the Python package \texttt{cosmoabc} \citep{ish15}, following \citet{kun20}. We choose $M$ = 1000 and $N$ = 200, and consider the algorithm converged when at least 20,000 draws are necessary to construct the final particle system. To prevent the reporting of an outlier as our result, we run five inferences and concatenated the posteriors.

\begin{figure*}
    \centering
    \includegraphics[width=\linewidth]{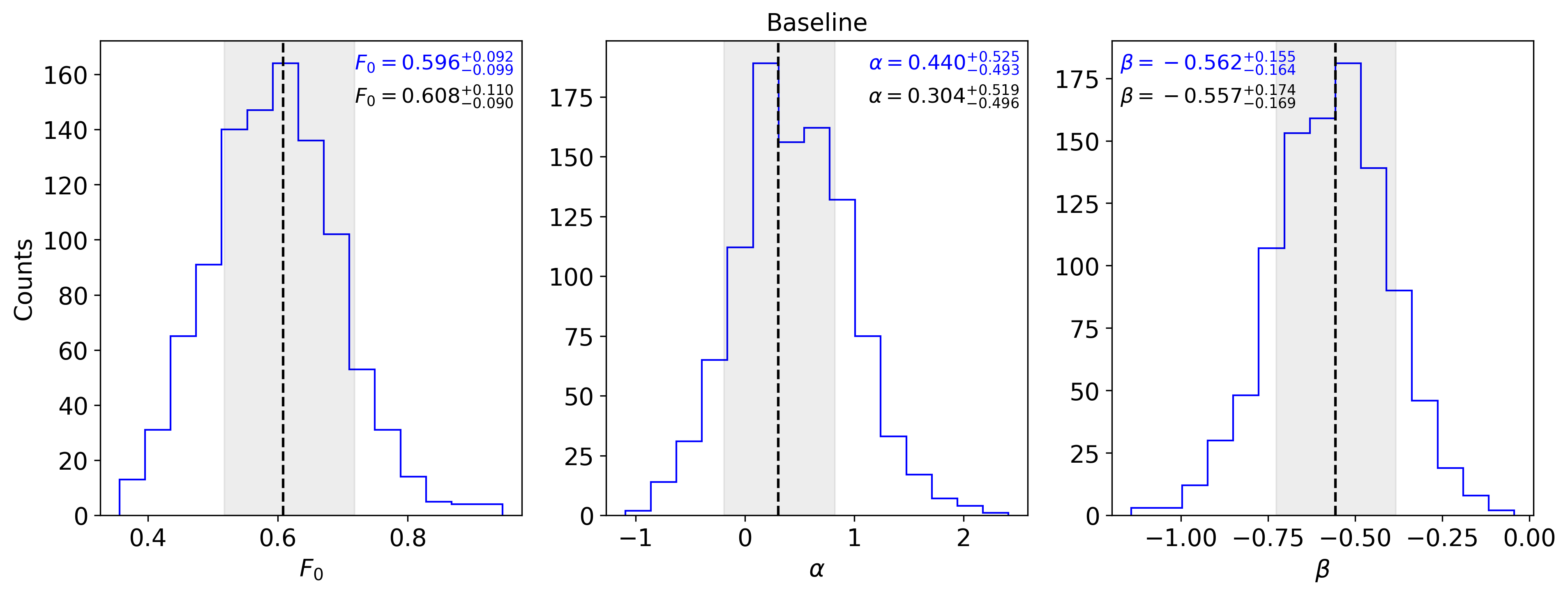}
    \includegraphics[width=\linewidth]{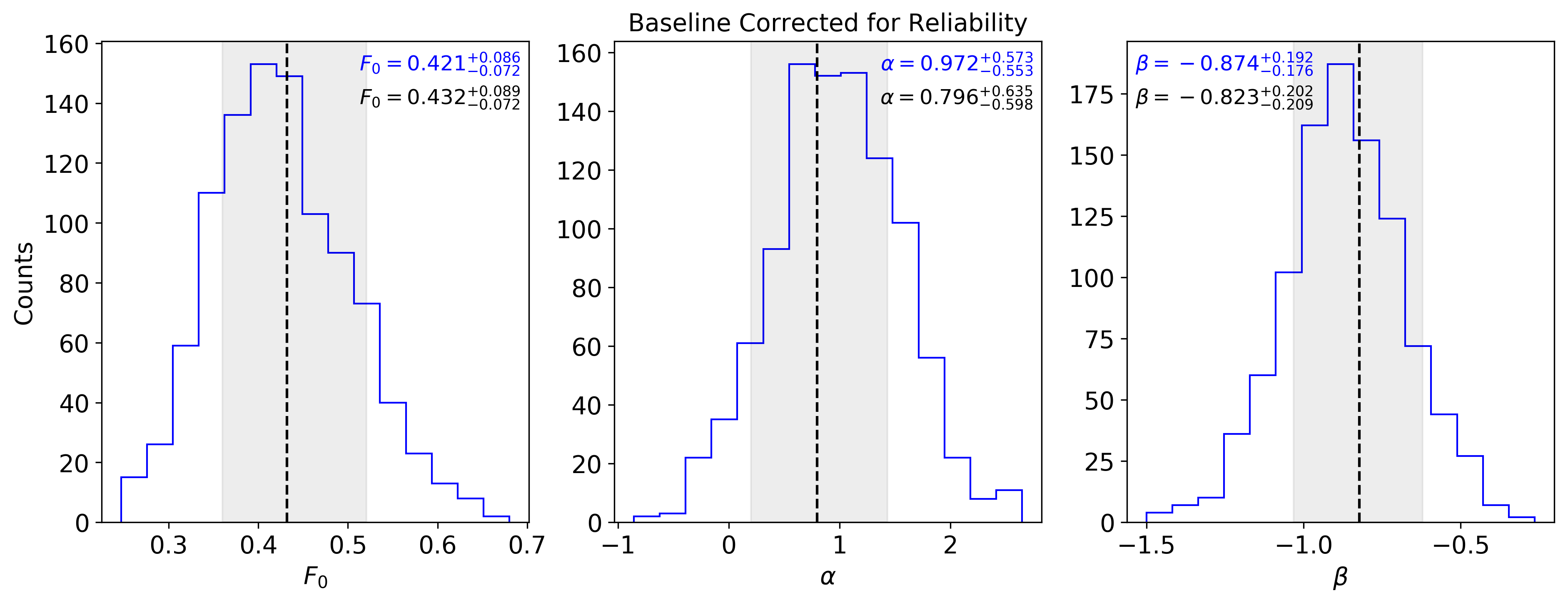}
    \caption{Occurrence rate results using ABC (blue histogram), compared to the baseline results from \citet{bry20} using the Poisson likelihood method (dotted black line, with grey shaded region representing 1$\sigma$ uncertainties).}
    \label{fig:results}
\end{figure*}

Fig. \ref{fig:results} compares our results to Table 1 of \citet{bry20}, who used Markov Chain Monte Carlo inference with a Poisson likelihood to find $F_{0} = 0.608_{-0.090}^{+0.110}$, $\alpha = 0.304_{-0.496}^{+0.519}$, and $\beta = {-0.557}_{-0.169}^{+0.174}$. We find $F_{0} = 0.596_{-0.099}^{+0.092}$, $\alpha = 0.440_{-0.493}^{+0.525}$, and $\beta = -0.562_{-0.164}^{+0.155}$, where the central value is the median and the uncertainties are the 16th and 84th percentiles of the ABC posterior.

\citet{bry20} also corrected for the reliability of the observed catalogue, recognizing that some planets may be astrophysical false positives or false alarms due to noise or systematics. They ran 100 inferences, probabilistically sampling from the observed planets according to their reliability each time, and concatenated the posteriors. They noted a significant drop in the exoplanet occurrence rate, finding $F_{0} = 0.432_{-0.072}^{+0.089}$, $\alpha=0.796_{-0.598}^{+0.635}$, and $\beta = -0.823_{-0.209}^{+0.202}$, demonstrating the importance of accounting for reliability. An advantage of ABC is that it is directly able to take reliability into account in a single run by weighting each observed planet's contribution to the distance, provided the distance can support weighted samples. After modifying the AD test and $\rho_{3}$ to accept weights, we found $F_{0} = 0.421_{-0.072}^{+0.086}$, $\alpha = 0.972_{-0.553}^{+0.573}$, and $\beta = -0.874_{-0.176}^{+0.192}$. Our parameters yield $\Gamma_{\oplus} \equiv d^{2}f/d\log{P}d\log{R_{p}}|_{P_{\oplus}, R_{\oplus}} = P_{\oplus}R_{\oplus}\lambda(P_{\oplus},R_{\oplus}) = 0.081_{-0.034}^{+0.052}$ and SAG13\footnote{https://exoplanets.nasa.gov/exep/exopag/sag/\#sag13} $\eta_{\oplus} = 0.108_{-0.046}^{+0.072}$.

These results agree within 1$\sigma$ of \citet{bry20}, indicating that the results are robust and not method-dependent.

%\detailtexcount{main}

\end{document}